\documentclass[twocolumn,pre,superscriptaddress]{revtex4}

\usepackage{dcolumn}
\usepackage{amsmath}

\usepackage{graphicx}

\newcommand{\half}{\tfrac12}
\newcommand{\etal}{{\it{}et~al.}}
\newcommand{\defn}{\textit}
\newcommand{\Ord}{\mathrm{O}}
\newcommand{\mat}{\mathbf}
\renewcommand{\vec}{\mathbf}

\begin{document}
\title{Multiway spectral community detection in networks}

\author{Xiao Zhang}
\affiliation{Department of Physics, University of Michigan, Ann Arbor,
  Michigan, USA}
\author{M. E. J. Newman}
\affiliation{Department of Physics, University of Michigan, Ann Arbor,
  Michigan, USA}
\affiliation{Center for the Study of Complex Systems, University of
Michigan, Ann Arbor, Michigan, USA}

\begin{abstract}
  One of the most widely used methods for community detection in networks
  is the maximization of the quality function known as modularity.  Of the
  many maximization techniques that have been used in this context, some of
  the most conceptually attractive are the spectral methods, which are
  based on the eigenvectors of the modularity matrix.  Spectral algorithms
  have, however, been limited by and large to the division of networks into
  only two or three communities, with divisions into more than three being
  achieved by repeated two-way division.  Here we present a spectral
  algorithm that can directly divide a network into any number of
  communities.  The algorithm makes use of a mapping from modularity
  maximization to a vector partitioning problem, combined with a fast
  heuristic for vector partitioning.  We compare the performance of this
  spectral algorithm with previous approaches and find it to give superior
  results, particularly in cases where community sizes are unbalanced.  We
  also give demonstrative applications of the algorithm to two real-world
  networks and find that it produces results in good agreement with
  expectations for the networks studied.
\end{abstract}

\maketitle

\section{Introduction}
\label{sec:intro}
Community detection, the division of the vertices of a network into groups
such that connections are dense within groups and sparser between them, has
been a topic of vigorous research, particularly within statistical physics,
for some years~\cite{Fortunato10}.  A broad range of different approaches
to the problem have been tried, but perhaps those in widest current use are
methods based on modularity maximization.  Modularity~\cite{NG04} is a
scalar objective function which assigns a numerical score to any division
of a network into communities, with higher scores being associated with
divisions that are better in the sense of having more edges within
communities and fewer between them.  Modularity maximization discovers good
divisions of a network by finding the ones that have the highest modularity
scores.  Unfortunately, the exhaustive numerical maximization of modularity
over all divisions of a network is known to be an NP-hard
task~\cite{Brandes07}, computationally tractable only for the very smallest
of networks, so we are forced to rely on approximate optimization
heuristics, a large number of which have been tried.  These include greedy
algorithms~\cite{Newman04a,CNM04}, simulated annealing~\cite{GSA04,MAD05},
extremal optimization~\cite{DA05}, genetic algorithms~\cite{SYHF09}, and
the widely used multiscale ``Louvain method'' of
Blondel~\etal~\cite{BGLL08}, which has been incorporated into a number of
common software packages.

In this paper we focus on another class of algorithms for modularity
maximization, the spectral algorithms, which are based on the examination
of the leading eigenvalues and eigenvectors of the so-called modularity
matrix~\cite{Newman06b}.  These methods are of interest for a number of
reasons.  First, they give high-quality results in practical situations
while also being fast, the eigenvalues and vectors normally being
calculated using the Lanczos method~\cite{Meyer00}, which is highly
efficient for the sparse matrices that typically arise in network problems.
Second, they are conceptually attractive, being based on well-understood
principles of linear algebra.  And third, they are, by contrast with most
other approaches, amenable to formal analysis, for instance using random
matrix theory~\cite{NN12}, allowing one to make precise statements about
their performance.

Spectral methods, however, do have their problems.  A primary one is that
there is no simple principled spectral algorithm for dividing a network
into an arbitrary number of communities.  Good algorithms exist for two-
and three-way divisions, and repeated two-way divisions can sometimes
produce good multiway divisions, but sometimes
not~\cite{Newman06b,Newman06c,RMP09}.  A better approach, proposed by White
and Smyth~\cite{WS05}, is to compute several leading eigenvectors of the
modularity matrix at once, represent them as points in a high-dimensional
space, and then cluster those points using a conventional data clustering
method---White and Smyth use $k$-means.  This method, which is analogous to
previous algorithms for the different but related problem of Laplacian
spectral graph partitioning~\cite{Elsner97,Fjallstrom98}, is attractive in
that it directly divides a network into the desired number of communities.
On the other hand, while the strong similarity between graph partitioning
and modularity maximization~\cite{Newman06c,Newman13b} makes it natural to
think that $k$-means would work in this situation, it is not clear what
quantity, if any, the algorithm of~\cite{WS05} is optimizing.  In
particular, the algorithm is not derived as an approximation to modularity
maximization, so there are no formal guarantees that it will indeed
maximize modularity, and in practice, as we show in this paper, there are
situations were it can fail badly.

In this paper, therefore, we introduce a different method for single-step,
multiway, spectral community detection.  Our method is not a generalization
of the previous two-way method, which is based on a relaxation of the
discrete modularity optimization problem to a continuous optimization that
can be solved by differentiation.  Instead the method is based on the
observation, made previously in~\cite{Newman06c}, that modularity
maximization is equivalent to a max-sum vector partitioning problem.  (A
similar equivalence for the graph partitioning problem was explored
in~\cite{AKY99,AY95}.)  We propose a simple heuristic for the rapid
solution of vector partitioning problems and apply it to the task in hand
to create an efficient multiway community detection algorithm.

\section{Spectral community detection and vector partitioning}
\label{sec:vector}
The modularity~$Q$ is a score assigned to a given division into any number
of communities of a given network, such that good divisions---those in
which most edges fall within communities and few edges fall between
them---get a high score and bad divisions a low one.  Formally, the
modularity is equal to the fraction of edges that fall within communities
minus the expected fraction if edges are placed at random~\cite{NG04}.
Consider an undirected, unweighted network of $n$ vertices and define an
adjacency matrix~$\mat{A}$ to represent the network structure with
elements~$A_{ij}=1$ if vertices~$i$ and~$j$ are connected by an edge and 0
otherwise.  Now consider a division of the vertices of this network
into~$k$ non-overlapping groups, labeled by integers~$1\ldots k$, and
define~$g_i$ to be the label of the group to which vertex~$i$ belongs.
Then the modularity is given by~\cite{CNM04}
\begin{equation}
Q = {1\over 2m} \sum_{ij} \biggl[ A_{ij} - {d_i d_j\over2m} \biggr]
  \delta_{g_i,g_j},
\label{eq:modularity0}
\end{equation}
where $d_i$ is the degree of vertex~$i$, $m$~is the total number of edges
in the network, and $\delta_{st}$ is the Kronecker delta.  The modularity
may be either positive or negative (or zero), with a maximum value of~$+1$.
Positive values indicate that the number of edges within groups is greater
than what one would expect by chance, and large positive values are
considered indicative of a good network division.

For convenience we also define the modularity matrix to be the symmetric
$n\times n$ matrix~$\mat{B}$ with elements
\begin{equation}
B_{ij} = A_{ij} - {d_i d_j\over 2m},
\label{eq:B}
\end{equation}
in terms of which the modularity~\eqref{eq:modularity0} can be written
\begin{equation}
Q = {1\over2m} \sum_{ij} B_{ij} \delta_{g_i,g_j}.
\label{eq:modularity1}
\end{equation}
Given that $\sum_i A_{ij} = d_j$ and $\sum_i d_i = 2m$, every row and
column of the modularity matrix must sum to zero:
\begin{equation}
\sum_i B_{ij} = \sum_i A_{ij} - \sum_i { d_i d_j \over 2m } = 0,
\label{eq:zerosum}
\end{equation}
which implies that the uniform vector $\vec{1} = (1,1,1,\dots)$ is an
eigenvector of the modularity matrix with eigenvalue zero, a result that
will be important shortly.

Now consider the problem of dividing a network with $n$ vertices into $k$
communities.  Since good divisions have high modularity scores and low
divisions low scores, we can find good divisions by maximizing modularity
over divisions.  Exact maximization is known to be very
slow~\cite{Brandes07}, so we turn instead to approximate methods.
Following~\cite{AY95,Newman06c}, we note that the delta function in
Eq.~\eqref{eq:modularity1} can be written as
\begin{equation}
\delta_{g_i,g_j} = \sum_{s=1}^k \delta_{s,g_i} \delta_{s,g_j},
\label{eq:delta}
\end{equation}
and since the modularity matrix is symmetric it can always be written as an
eigenvector decomposition
\begin{equation}
B_{ij} = \sum_{l=1}^n \lambda_l U_{il} U_{jl},
\label{eq:eigen}
\end{equation}
where $\lambda_l$ is an eigenvalue of~$\mat{B}$ and $U_{il}$ is an element
of the orthogonal matrix~$\mat{U}$ whose columns are the corresponding
eigenvectors.  Without loss of generality, we will assume that the
eigenvalues are numbered in decreasing order: $\lambda_1 \ge \lambda_2 \ge
\dots \ge \lambda_n$.  Combining Eqs.~\eqref{eq:modularity1},
\eqref{eq:delta}, and~\eqref{eq:eigen}, we now have
\begin{align}
Q &= {1\over2m} \sum_{ij} \sum_{l=1}^n \lambda_l U_{il} U_{jl}
      \sum_s \delta_{s,g_i} \delta_{s,g_j} \nonumber\\
  &= {1\over2m} \sum_{l=1}^n \lambda_l
     \sum_s \biggl[ \sum_i U_{il} \delta_{s,g_i} \biggr]^2.
\end{align}

We observe that (apart from the uninteresting leading constant) this is a
sum over eigenvalues~$\lambda_l$ times the nonnegative quantities $\sum_r
\bigl[ \sum_s U_{il} \delta_{s,g_i} \bigr]^2$, so the largest (most
positive) contributions to the modularity are typically made by the terms
corresponding to the most positive eigenvalues.  A standard approximation,
used in essentially all spectral algorithms, is, instead of maximizing the
entire sum, to maximize only these largest terms, neglecting the others.
That is, we approximate the modularity by
\begin{equation}
Q = {1\over2m} \sum_{l=1}^p \lambda_l
     \sum_s \biggl[ \sum_i U_{il} \delta_{s,g_i} \biggr]^2.
\label{eq:modularity2}
\end{equation}
for some integer~$p<n$.  At a minimum, we maximize only those terms
corresponding to positive values of~$\lambda_l$.  (Maximizing ones
corresponding to negative~$\lambda_l$ would reduce, not increase, the
modularity.)  In effect, we are making a rank-$p$ approximation to the
modularity matrix, based on its leading $p$ eigenvectors, then calculating
the modularity using that approximation rather than the true modularity
matrix.

Noting that all $\lambda_l$ in Eq.~\eqref{eq:modularity2} are now positive,
we can rewrite the equation as
\begin{equation}
Q = {1\over2m} \sum_{s=1}^k \sum_{l=1}^p
    \biggl[ \sum_i \sqrt{\lambda_l} U_{il} \delta_{s,g_i} \biggr]^2.
\label{eq:modularity3}
\end{equation}
We define a set of $n$ $p$-dimensional \defn{vertex vectors}~$\vec{r}_i$
with elements
\begin{equation}
\bigl[ \vec{r}_i \bigr]_l = \sqrt{\lambda_l} U_{il},
\label{eq:r}
\end{equation}
in terms of which the modularity is
\begin{equation}
Q = {1\over2m} \sum_{s=1}^k \sum_{l=1}^p
    \biggl[ \sum_{i\in s} \bigl[ \vec{r}_i \bigr]_l \biggr]^2
  = {1\over2m} \sum_{s=1}^k \biggl| \sum_{i\in s} \vec{r}_i \biggr|^2,
\label{eq:modularity4}
\end{equation}
where the notation~$i\in s$ denotes that vertex~$i$ is in group~$s$.

In other words, we assign to each vertex a vector~$\vec{r}_i$, which can be
calculated solely from the structure of the network (since it is expressed
in terms of the eigenvalues and eigenvectors of the modularity matrix) and
hence is constant throughout the optimization procedure.  Then the
modularity of a division of the network into groups is given (apart from
the leading constant~$1/2m$) as a sum of contributions, one from each
group~$s$, equal to the square of the sum of the vectors for the vertices
in that group.  Our goal is to find the division that maximizes this
modularity.

Generically, problems of this kind are called \defn{max-sum vector
  partitioning} problems, or just vector partitioning for short.  In the
following section we propose a heuristic algorithm to rapidly solve vector
partitioning problems and show how it can be applied to perform efficient
multiway spectral community detection in arbitrary networks.

We have not yet said what the value should be of the constant~$p$ that
specifies the rank at which we approximate the modularity matrix in
Eq.~\eqref{eq:modularity2}.  We have said that $p$ should be no greater
than the number of positive eigenvalues of the modularity matrix.  On the
other hand, as shown in~\cite{Newman06c}, if $p$ is less than $k-1$ then
the division of the network with maximum modularity always has less than
$k$ communities, since there will be at least one pair of communities whose
amalgamation into a single community will increase the modularity.  Thus
$p$ should be greater than or equal to $k-1$.  In all of the calculations
presented in this paper we make the minimal choice $p=k-1$, which gives the
fastest algorithm and in most cases gives excellent results.  However, it
is worth bearing in mind that larger values of $p$ are possible and, in
principle, give more accurate approximations to the true value of the
modularity.

\section{Vector partitioning algorithm}
\label{sec:algo}
Vector partitioning is computationally easier than many optimization tasks.
In particular, it is solvable in polynomial, rather than exponential time.
A general $k$-way partitioning of $n$ different $p$-dimensional vectors can
be solved exactly in time $\Ord(n^{p(k-1)-1})$~\cite{OS01}.  Thus if we use
the leading two eigenvectors of the modularity matrix to divide a network
into two communities the calculation can be done in time $\Ord(n)$, as
shown previously in~\cite{Newman06c}.  However the running time quickly
becomes less tractable for larger numbers of communities.  As discussed
above, for a division of a network into $k$ communities we must use at
least $k-1$ eigenvectors, which gives a running time $\Ord{(n^{k^2-2k})}$.
Even for just three communities this gives~$\Ord(n^3)$, which is practical
only for rather small networks, and for four communities it
gives~$\Ord(n^8)$ which is entirely impractical.  For applications to
realistically large networks with $k>2$, therefore, we must abandon exact
solution of the problem and look for faster approximate methods.

Previous approaches to vector partitioning include that of
Wang~\etal~\cite{WSO08}, who suggest dividing the space of vectors into
octants (or their generalization in higher dimensions) and looking through
all $2^{k-1}$ of them to find the $k$ octants that contain the largest
numbers of vectors.  Then we use these as an initial coarse division and
assign the remaining vectors to these groups by brute-force optimization.
This method works reasonably well for small values of~$k$ but is not ideal
as $k$ becomes larger because the number of octants increases exponentially
with~$k$.  Richardson~\etal~\cite{RMP09} proposed a divide-and-conquer
method that works by splitting the space into octants again, but then
splitting these into smaller wedges, and repeating until further
subdivision gives no improvement.  This method works well for the $k=3$
case with two eigenvectors but does not generalize well to higher~$k$.
Alpert and Yao~\cite{AY95} proposed a greedy algorithm that works for any
value of $k$ by adding vectors one by one to the set to be partitioned,
with vectors of larger magnitude being added first (on the grounds that
these contribute most to the sums in Eq.~\eqref{eq:modularity4}).  This
method works well when the largest magnitude vectors are distributed evenly
among the final groups, but more poorly when they are concentrated in a few
groups.  Unfortunately, as we show in Section~\ref{sec:synthetic}, when
network communities are of unequal sizes the largest vertex vectors do
indeed tend to be concentrated in a few groups and the method
of~\cite{AY95} works less well.
 
Here we introduce an alternative and well-motivated heuristic for finding
the solution to vector partitioning problems for general values of~$k$.
The algorithm is analogous to the $k$-means algorithm for the standard data
partitioning problem.  The $k$-means method is an algorithm for
partitioning a set of data points in any number of dimensions into $k$
clusters in which we start by choosing $k$ index locations or centroids in
the space.  These could be chosen in several ways: entirely at random, at
random from among the set of data points, or (most commonly) as the
centroids of some initial approximate partition of the data.  Once these
are chosen, we compute the distance from each data point to each of the $k$
centroids and divide the data points into $k$ groups according to which
they are closest to.  Then we compute the $k$ centroids of these new
groups, replace the old centroids with the new ones, and repeat.  The
process continues until the centroids stop changing.

Our algorithm adopts a similar idea for vector partitioning, with points
being replaced by vectors and distances replaced by vector inner products.
We start by choosing an initial set of $k$ \defn{group
  vectors}~$\vec{R}_s$, one for each group or community~$s$, then we assign
each of our vertex vectors~$\vec{r}_i$ to one of the groups according to
which group vector it is closest to, in a sense we will define in a moment.
Then we calculate new group vectors for each community from these
assignments and repeat.  The new group vectors are calculated simply as the
sums of the vertex vectors in each group:
\begin{equation}
\vec{R}_s = \sum_{i\in s} \vec{r}_i,
\label{eq:R}
\end{equation}
so that the modularity, Eq.~\eqref{eq:modularity4}, is equal to
\begin{equation}
Q = {1\over2m} \sum_s \bigl|\vec{R}_s\bigr|^2.
\end{equation}
We observe the following property of this modularity.  Suppose we move a
vertex~$i$ from one community~$s$ to another~$t$.  Let $\vec{R}_s$
and~$\vec{R}_t$ represent the group vectors of the two communities
excluding the contribution from vertex~$i$.  Then, before the move, the
group vectors of the communities are $\vec{R}_s+\vec{r}_i$ and $\vec{R}_t$,
and after the move they are $\vec{R}_s$ and $\vec{R}_t+\vec{r}_i$.  All
other communities remain unchanged in the meantime and hence the
change~$\Delta Q$ in the modularity upon moving vertex~$i$ is
\begin{align}
\Delta Q &= {1\over2m} \bigl[ |\vec{R}_s|^2 + |\vec{R}_t+\vec{r}_i|^2
           - |\vec{R}_s+\vec{r}_i|^2 - |\vec{R}_t|^2 \bigr] \nonumber\\
  &= {1\over m} \bigl[ \vec{R}_t^T\vec{r}_i - \vec{R}_s^T\vec{r}_i \bigr].
\label{eq:RsRt}
\end{align}
Thus the modularity will either increase or decrease depending on which is
the larger of the two inner products $\vec{R}_t^T\vec{r}_i$ and
$\vec{R}_s^T\vec{r}_i$.  Or, to put that another way, in order to maximize
the modularity we should assign to vertex~$i$ to the community whose group
vector has the largest inner product with~$\vec{r}_i$.

This then defines our equivalent of ``distance'' for our $k$-means style
vector partitioning algorithm.  Given a set of group vectors~$\vec{R}_s$,
we calculate the inner product~$\vec{R}_s^T\vec{r}_i$ between $\vec{r}_i$
and every group vector and then assign vertex~$i$ to the community with the
highest inner product.

Note, however, that the group vectors~$\vec{R}_s$ and~$\vec{R}_t$ appearing
in Eq.~\eqref{eq:RsRt} are defined \emph{excluding}~$\vec{r}_i$ itself.  To
be correct, therefore, we should do the same thing in our partitioning
algorithm.  For every vertex vector~$\vec{r}_i$ there will be one group
vector~$\vec{R}_s$ that contains that vertex vector (in the sense of
Eq.~\eqref{eq:R}) and before calculating the inner product for that group
we should subtract~$\vec{r}_i$ from the group vector.  In practice this
subtraction typically makes little difference when the network is
large---the subtraction or not of a single vertex from a large group is not
going to change the results much.  In many cases, therefore, one can omit
the subtraction step.  On the other hand, the algorithm is not
significantly slower with the subtraction, so one could also argue for its
inclusion, purely on grounds of correctness.  We do include it in the
calculations of this paper, but in the end it makes little difference to
the results.

\begin{figure}
\begin{center}
\includegraphics[width=\columnwidth,clip=true]{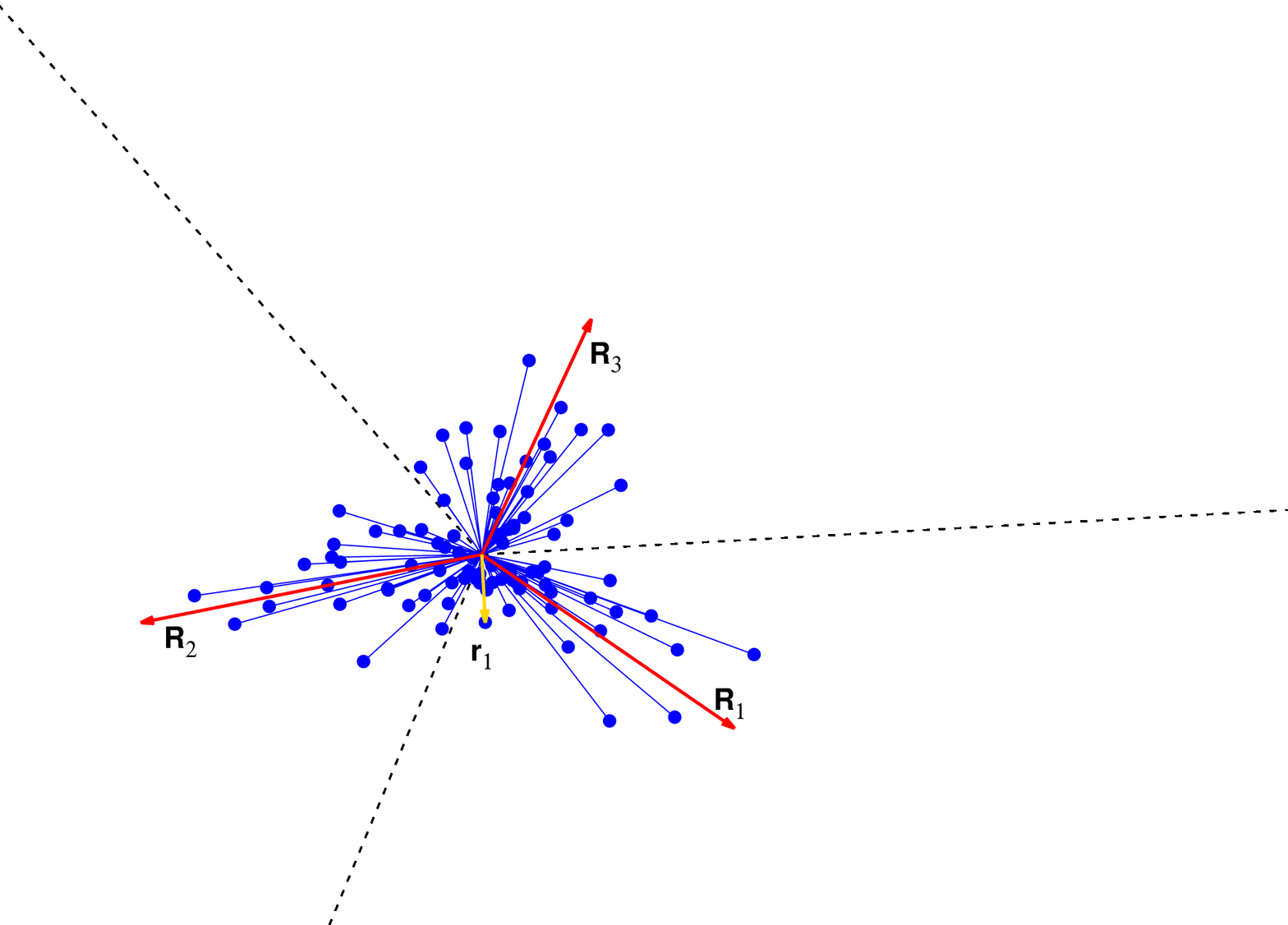}
\end{center}
\caption{Depiction of the operation of our vector partitioning heuristic
  for, in this case, a set of two-dimensional vectors being divided into
  three groups.  The blue lines and dots denote the individual vectors.
  The red lines are the group vectors.  (The magnitudes of the group
  vectors have been rescaled to fit into the figure---normally they would
  be much larger, since they are the sums of the individual vectors in each
  group.)  The dashed lines indicate the borders between communities, which
  are determined both by the angles and relative magnitudes of the group
  vectors.  For example, the vector labeled~$\vec{r}_1$ will be assigned to
  group~1 in this case, because it has its largest inner product
  with~$\vec{R}_1$.}
\label{fig:illustrate}
\end{figure}

Our complete vector partitioning algorithm is
the following:
\begin{enumerate}
\item Choose an initial set of group vectors~$\vec{R}_s$, one for each of
  the $k$ communities.
\item Compute the inner product $\vec{R}_s^T\vec{r}_i$ for all vertices~$i$
  and all communities~$s$, or $(\vec{R}_s-\vec{r}_i)^T\vec{r}_i$ if
  vertex~$i$ is currently assigned to group~$s$.
\item Assign each vertex to the community with which it has the highest (most
  positive) inner product.
\item Update the group vectors using the definition of Eq.~\eqref{eq:R}.
\item Repeat from step 2 until the group vectors stop changing.  (One could
  also halt when the changes become negligible or after some maximum number
  of iterations, just as some $k$-means implementations also do.)
\end{enumerate}
See Fig.~\ref{fig:illustrate} for an illustration of the working of the
algorithm.

We still need to decide how our the initial group vectors should be chosen.
In the simplest case we might just choose them to be of equal magnitude and
point in random directions.  However, if there is community structure in
the network then we expect the vertex vectors to be clustered, pointing in
a small number of directions, with no or few vectors pointing in the
remaining directions.  It makes little sense to pick initial group vectors
pointing in directions well away from where the clusters lie, so in
practice we have found that, rather than giving the group vectors random
directions, we can get good results by picking them randomly from among the
vertex vectors themselves.  This ensures that, if most vectors point in a
few directions, we will be likely to choose initial group vectors that also
point in those directions.

Note in fact that we need only pick $k-1$ of the $k$ group vectors in this
fashion, the final vector being fixed by the fact that the group vectors
sum to zero.  To see this, recall that the uniform vector $\vec{1} =
(1,1,1,\ldots)$ is always an eigenvector of the modularity matrix, which
implies that the elements of all other eigenvectors---i.e.,~the columns of
the orthogonal matrix~$\mat{U}$---must sum to zero (since they must be
orthogonal to the uniform vector).  Then the definition of Eq.~\eqref{eq:r}
implies that
\begin{equation}
\sum_{i=1}^n \bigl[ \vec{r}_i\bigr]_l
  = \sqrt{\lambda_l} \sum_{i=1}^n U_{il} = 0,
\end{equation}
and hence
\begin{equation}
\sum_{i=1}^n \vec{r}_i = 0,
\label{eq:cofm}
\end{equation}
and
\begin{equation}
\sum_s \vec{R}_s = \sum_s \sum_{i\in s} \vec{r}_i
  = \sum_{i=1}^n \vec{r}_i =0.
\label{eq:Rconstraint}
\end{equation}
Thus, once we have chosen $k-1$ of the group vectors randomly, the final
one is fixed to be equal to minus the sum of the rest.

Since there is a random element in the initialization of our algorithm, its
result is not always guaranteed to be the same, even when applied to the
same network with the same parameter values; it may give different results
for the modularity on different runs.  In applications, therefore, we
typically do several runs of the algorithm with different initial
conditions, choosing from among the results the community division that
gives the highest value of the modularity.

\section{Applications}
\label{sec:examples}
In this section we give example applications of our method, first to
computer-generated test networks and then to two real-world examples.

\subsection{Synthetic networks}
\label{sec:synthetic}
For our first tests of the method we look at a set of computer-generated
(``synthetic'') benchmark networks that contain known community structure.
Our goal is to see whether, and how accurately, the algorithm can recover
that structure.  In our tests we make use of networks generated using the
\defn{degree-corrected stochastic block model}~\cite{KN11a}.  The
\defn{stochastic block model} (not degree-corrected) is a generative model
of community-structured networks whose origins go back to the
1980s~\cite{HLL83,CL09}.  Vertices are divided into groups and edges are
placed between pairs independently at random with
probabilities~$\omega_{st}$ that depend only on the groups~$s,t$ that the
vertices belong to.  If the diagonal probabilities~$\omega_{ss}$ are larger
than the off-diagonal ones, then the network will display classic
``assortative'' community structure with more connections within groups
than between them.  The stochastic block model is unrealistic, however, in
generating a Poisson distribution of vertex degrees, which is quite
different from the highly right-skewed distributions commonly seen in real
networks.  The degree-corrected block model remedies this problem by fixing
the (expected) degrees of the vertices at any values we choose.  In this
model edges are placed independently between pairs of vertices~$i,j$ with
probability $d_i d_j \omega_{st}$, where $d_i$ is the desired degree of
vertex~$i$.  For a detailed discussion see~\cite{KN11a}.

Our tests consist of generating a number of networks using the
degree-corrected block model, analyzing them using our algorithm, then
comparing the communities found with those planted in the networks in the
first place.  To quantify the similarity of the two sets of communities,
planted and detected, we make use of a standard measure, the
\defn{normalized mutual information} or NMI~\cite{DDDA05,Meila07}.  The
(unnormalized) mutual information of two sets~$X,Y$ of numbers or
measurements is defined to be
\begin{equation}
I(X;Y) = \sum_{x \in X} \sum_{y \in Y} p(x,y) \log{p(x,y) \over p(x)p(y)},
\label{eq:MI}
\end{equation}
where $p(x,y)$ is the joint probability or frequency of $x$ and~$y$ within
the data set and $p(x)$, $p(y)$ are their marginal probabilities.  The
mutual information measures how much you learn about one of the two sets of
measurements by knowing the other.  If $X$ and $Y$ are uncorrelated then
each tells you nothing about the other and the mutual information is zero.
If they are perfectly correlated then each tells you everything about the
other and the mutual information takes its maximum value, which is equal to
all of the information that either set contains, which is simply the
entropy, $H(X)$~or~$H(Y)$, of the set.

\begin{figure*}[ht]
\begin{center}
\includegraphics[width=2\columnwidth]{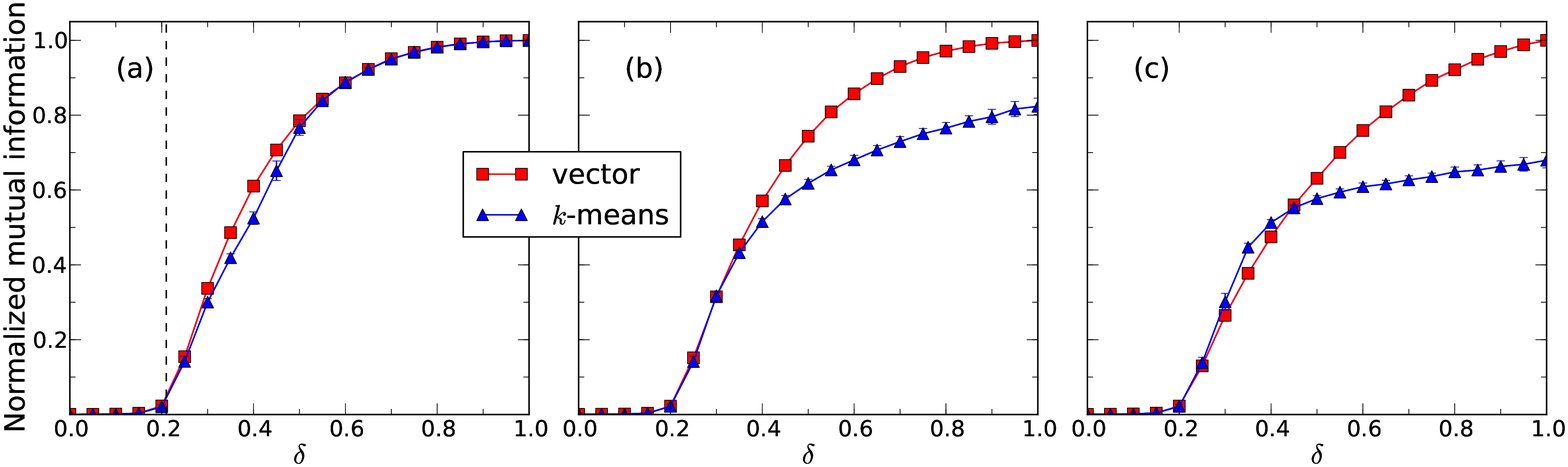}
\end{center}
\caption{Normalized mutual information as a function of the parameter
  $\delta$ for communities detected in randomly generated test networks
  using the vector partitioning algorithm of this paper (red squares) and
  the $k$-means method of Ref.~\cite{WS05} (blue triangles).  The networks
  consist of $n=3600$ vertices each, divided into three communities thus:
  (a)~equally sized communities of 1200 vertices each; (b)~communities of
  sizes 1800, 1200, and~600; (c)~communities of sizes 2400, 900, and~300.
  Each data point is an average of 100 networks.  The vertical dashed line
  in panel~(a) indicates the position of the detectability threshold below
  which all methods must fail~\cite{DKMZ11a}.}
\label{fig:sbm}
\end{figure*}

Having the maximum value of the mutual information be equal to the entropy
is in some ways inconvenient, since we don't know in advance what that
value will be.  So commonly one normalizes the mutual information by
dividing by the mean of the entropies of the two sets, thus:
\begin{equation}
\textrm{NMI}(X;Y) = {I(X;Y)\over\half[H(X) + H(Y)]}.
\label{eq:nmi}
\end{equation}
This normalized value falls in the interval from zero to one, with
uncorrelated variables giving zero and perfect correlation giving one.

The NMI is commonly used to quantify the match between two clusterings of
the vertices of a network.  In the present case, the original assignments
of vertices to groups in the block model (the ``planted communities'') are
used as one set of measurements~$X$ and the assignments found by our
algorithm (the ``inferred communities'') are the other~$Y$.  An NMI of 1
denotes perfect recovery of the planted partition; an NMI of 0 indicates
complete failure.

In the tests presented here we use networks of $n=3600$ vertices divided
into $k=3$ communities and with two different (expected) degrees: half the
vertices in each group have degree~10 and the other half have degree~30.
The parameters~$\omega_{st}$ are varied in order to tune the difficulty of
the community detection according to
\begin{equation}
 \omega_{st} = (1-\delta) \omega_{st}^{\text{random}}
               + \delta \omega_{st}^{\textrm{planted}},
\label{eq:omega}
\end{equation}
where $\delta$ is a parameter that varies from zero to one and
\begin{equation}
\omega_{st}^{\textrm{random}} = {1\over2m},
\qquad
\omega_{st}^{\textrm{planted}} = {\delta_{st}\over\sum_{i\in s} d_i},
\end{equation}
with $m$ being the total number of edges in the network, as previously.
With this choice, the parameter~$\delta$ tunes the edge probabilities from
a value of $d_i d_j/2m$ when $\delta=0$, which corresponds to a purely
random edge distribution with no community structure at all (the so-called
configuration model~\cite{MR95,NSW01,CL02a}) to a value of $d_i
d_j/\sum_{i\in s} d_i$ within each group~$s$ and zero between groups
when~$\delta=1$---effectively three separate, unconnected configuration
models, one for each group, which is the strongest form of community
structure one could have.  This choice of $\omega_{st}$ also has the nice
property that the expected fraction of within-group edges that a vertex has
is the same for all vertices.

We have tested our algorithm on these networks using two eigenvectors to
define the vertex vectors~$\vec{r}_i$ (the minimum viable number).  The
results are shown as a function of the parameter~$\delta$ in
Fig.~\ref{fig:sbm}, along with results for the same networks analyzed by
clustering the vertex vectors using the $k$-means algorithm of
Ref.~\cite{WS05}.

As $\delta\to1$ the community structure in the network becomes strong and
any reasonable algorithm should be able to detect it.  As we approach this
limit our algorithm assigns 100\% of vertices to their correct communities
and the NMI approaches one.  Conversely as $\delta\to0$ the community
structure in the network vanishes and neither algorithm should detect
anything, so NMI approaches zero.  Furthermore, it is known that there is a
critical strength of the structure---which translates to a critical value
of our parameter~$\delta$---below which the structure is so weak that no
algorithm can detect it~\cite{DKMZ11a}.  This ``detectability threshold''
is marked in Fig.~\ref{fig:sbm}a with a vertical dashed line.  Above this
point it should be possible to detect the communities, albeit with a
certain error rate, and indeed we see that both algorithms achieve a
nonzero NMI in this region.

\begin{figure}
\begin{center}
\includegraphics[width=8cm,clip=true]{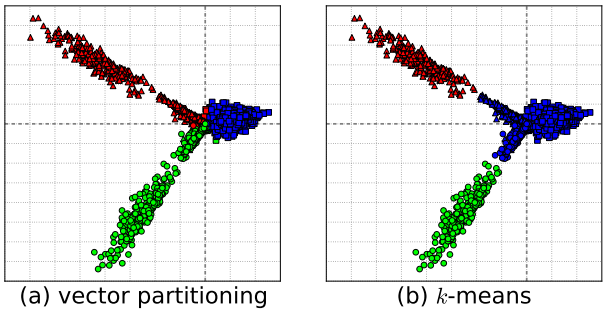}
\end{center}
\caption{Illustration of the division of a synthetic three-group network
  using (a)~the algorithm of this paper and (b)~the $k$-means algorithm
  of~\cite{WS05}.  Shapes indicate the planted communities while colors
  indicate the communities found by the two algorithms.  Observe how the
  $k$-means results assign a good portion of vertices belonging to the red
  and green communities incorrectly to the blue one, while the vector
  partitioning approach does not have this problem.  The network in this
  case has $n=4000$ vertices with communities of size 3000, 500, and~500.}
\label{fig:compare}
\end{figure}

As the figure shows, the vector partitioning algorithm does as well or
better than $k$-means in almost all cases.  In panel~(a) the three
communities in the network have equal sizes, and in this case the two
algorithms perform comparably, there being only a small range of parameter
values in the middle of the plot where vector partitioning outperforms
$k$-means by a narrow margin.  In panels~(b) and~(c) the communities have
unequal sizes---moderately so in (b) and highly in~(c)---and in these cases
vector partitioning does significantly better than $k$-means.  Indeed for
unequal group sizes the $k$-means algorithm fails to achieve perfect
community classification ($\textrm{NMI} = 1$) even in the limit where
$\delta=1$.  The reason for this is illustrated in Fig.~\ref{fig:compare},
which shows a scatter plot of the vertex vectors for an illustrative
example network along with the communities into which each algorithm
divides the vertices (shown by the colors).  As the figure shows, when the
groups are unequal in size the largest group is closer to the origin than
the smaller ones---necessarily so since the centroid of the vertex vectors
lies at the origin (Eq.~\eqref{eq:cofm}).  This tends to throw off the
$k$-means algorithm, which by definition splits the points into groups of
roughly equal spatial extent.  The vector partitioning method, which is
(correctly) sensitive only to the direction and not the magnitude of the
vertex vectors, has no such problems.

\subsection{Real-world examples}
\label{sec:realworld}

\begin{figure}
\begin{center}
\includegraphics[width=8cm,clip=true]{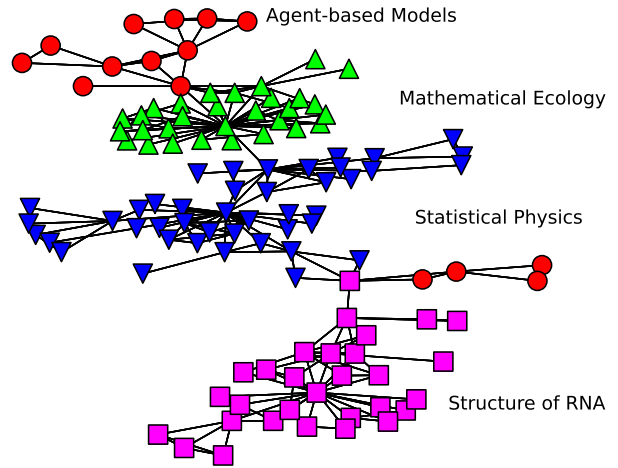}
\end{center}
\caption{Four-way division into communities of a collaboration network of
  scientists at the Santa Fe Institute.  Different colors and shapes
  indicate the communities discovered by the vector partitioning algorithm
  of this paper.  The communities split roughly along lines of research
  topic.}
\label{fig:santafe}
\end{figure}

\begin{figure*}
\begin{center}
\includegraphics[width=16cm,clip=true]{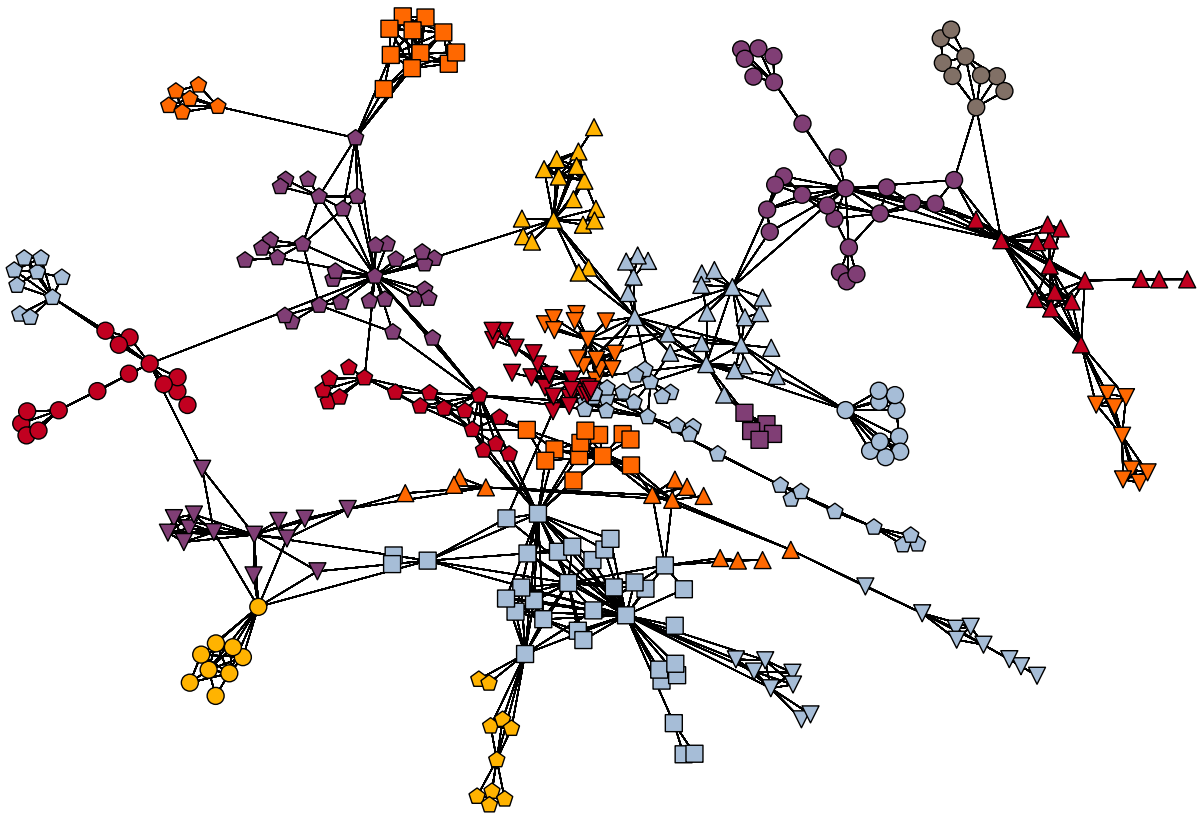}
\end{center}
\caption{The 21 communities found in a collaboration network of network
  scientists using the algorithm proposed in this paper.}
\label{fig:netsci}
\end{figure*}

Our next two example applications are to real-world networks, two
collaboration networks among scientists.  The first, taken from
Ref.~\cite{GN02}, represents scientists working at the Santa Fe Institute,
an interdisciplinary research institute in New Mexico.  The vertices in the
network represent the scientists and the edges indicate that two scientists
coauthored a paper together at least once.  The network is small enough to
allow straightforward visualization of our results and is interesting in
that the scientists it represents, in keeping with the interdisciplinary
mission of the institute, come from a range of different research fields,
in this case statistical physics, mathematical ecology, RNA structure, and
agent-based modeling.  It is plausible that the communities in the network
might reflect these subject areas.

Figure~\ref{fig:santafe} shows the result of a four-way community division
of this network using vertex vectors constructed from the first three
eigenvectors of the modularity matrix.  Overall the results mirror our
expectations, with the four subject areas corresponding roughly to the four
communities found by the method.  We note, however, that there are also
four vertices in the middle-right of the figure that are clearly
misclassified as being in the ``agent-based models'' group when they would
be more plausibly placed in the ``structure of RNA'' group.  This
illustrates a potential weakness of the algorithm: the defining feature of
these vertices is that their vertex vectors have very small magnitude,
meaning that they do not strongly belong to any group.  For such vertices
even a small error---such as that introduced by making our low-rank
approximation to the true modularity matrix---can alter the direction of
the vertex vector substantially and hence move a vertex to a different
group.  Problems like this are, in fact, common to many spectral algorithms
and are typically handled by combining the algorithm with a subsequent
iterative refinement or ``fine tuning'' step, in which individual vertices
or small sets of vertices are moved from group to group in an effort to
improve the value of the modularity~\cite{Newman06b,RMP09}.  The spectral
algorithm is good at determining the ``big picture,'' rapidly doing an
overall division of the network into broad groups of vertices; the
subsequent fine tuning tidies up the remaining details.  Based on the
results we see here, our algorithm might be a good candidate for
combination with a fine tuning step of this kind.

Our second real-world example is a collaboration network of scientists
working in the field of network science itself and is taken from
Ref.~\cite{Newman06c}.  Apart from being rather larger than the Santa Fe
Institute network, at 379 scientists, this network also differs in that all
its members are, ostensibly at least, studying the same subject, so there
is no obvious ``ground truth'' for the communities as there was in the
previous example, or even for how many communities there should be.
Choosing the number of communities into which a network should be divided
is a deep problem in its own right, and one that is not completely solved.
Here, however, we simply borrow a technique from the literature and
estimate the number of communities in the network by counting the real
eigenvalues of the so-called non-backtracking matrix that are greater than
the largest real part among the complex eigenvalues.  (For a discussion of
why this is a good heuristic, see~\cite{Krzakala13}.)  In the present case
this suggests that there should be 26 communities in the network, so we
choose $k=26$ for our community detection algorithm and construct the
vertex vectors from the leading 25 eigenvectors of the modularity matrix.
The results are shown in Fig.~\ref{fig:netsci}.  In fact, in this case we
find that the algorithm does not make use of all 26 communities---the
figure contains only 21.  Nonetheless, the algorithm has succeeded in
finding a good division in terms of modularity: the modularity value is $Q
= 0.83$, comparable to the value given for example in~\cite{RMP09} for the
same network.  We note, however, that, as is typical for larger values
of~$k$, the algorithm finds a range of different divisions of the network
in different runs that all have competitive modularity.  The existence of
competing good community divisions in the same network is a well-known
phenomenon and has been previously discussed for instance by
Good~\etal~\cite{GDC10}.

\section{Conclusions}
In this paper we have described a mapping of a multiway spectral community
detection method onto a vector partitioning problem and proposed a simple
heuristic algorithm for vector partitioning that returns good results in
this application.  We have tested our method on computer-generated
benchmark networks, comparing it with a competing spectral algorithm that
makes use of $k$-means clustering, and find our method to give superior
performance, particularly in cases where the sizes of the communities are
unequal.  We have also given two example applications of our method to
real-world networks.

There remain a number of open questions not answered in this paper.
Although the algorithm we propose is simple and efficient, it is only
approximate and we have no formal results on its expected performance.  The
algorithm also assumes we have prior knowledge of the number of communities
in the network, where in reality this is not usually the case.  Determining
the number of communities in a network is an interesting open problem.
Finally, as we (and others) have pointed out, the best community detection
methods are typically hybrids of two or more elementary methods.  It would
be interesting to see how the vector partitioning algorithm we propose
works in combination with other methods.  These problems, however, we leave
for future work.

\begin{acknowledgements}
  The authors thank Maria Riolo and Raj Rao Nadakuditi for helpful
  conversations.  This research was funded in part by the US National
  Science Foundation under grants DMS--1107796 and DMS--1407207.
\end{acknowledgements}

\end{document}